\documentclass[10pt,letterpaper]{article}
\usepackage[top=0.85in,left=2.75in,footskip=0.75in,marginparwidth=2in]{geometry}

\newcommand{\toolname}{\textit{line\_explorer}}

\usepackage[utf8]{inputenc}

\usepackage{cite}

\usepackage{nameref,hyperref}

\usepackage[right]{lineno}

\usepackage{microtype}
\DisableLigatures[f]{encoding = *, family = * }

\raggedright
\usepackage{changepage}

\usepackage[aboveskip=1pt,labelfont=bf,labelsep=period,singlelinecheck=off]{caption}

\makeatletter
\renewcommand{\@biblabel}[1]{\quad#1.}
\makeatother

\usepackage{lastpage,fancyhdr,graphicx}
\usepackage{epstopdf}
\fancyheadoffset[L]{2.25in}
\fancyfootoffset[L]{2.25in}

\usepackage{color}

\definecolor{Gray}{gray}{.25}

\usepackage{graphicx}

\usepackage{sidecap}

\usepackage{wrapfig}
\usepackage[pscoord]{eso-pic}
\usepackage[fulladjust]{marginnote}
\reversemarginpar

\begin{document}
\vspace*{0.35in}

\begin{flushleft}
{\Large
\textbf\newline{Assessing the Usability of a Novel System for Programming Education}
}
\newline
\\
Giovanni Vincenti\textsuperscript{1,*},
Scott Hilberg\textsuperscript{2},
James Braman\textsuperscript{3},
Michael Satzinger\textsuperscript{1},
Lily Cao\textsuperscript{2}
\\
\bigskip
\bf{1} University of Baltimore
\\
\bf{2} Towson University
\\
\bf{3} Community College of Baltimore County
\\
\bigskip
* gvincenti@ubalt.edu

\end{flushleft}

\section*{Abstract}
The authors present the results of a simple usability test performed on \toolname, an innovative tool aimed at letting students explore programming. The system offers an interactive environment where students can learn, review, and practice programming independently or through step-by-step instruction. Students in Information Technology, Computer Science, and Information Systems were surveyed. The findings show that students have interest in this tool, whereas some groups find this tool more interesting and useful. The findings will help refine the user interface for the next phase of testing which include changes for simplicity, usability and expanded topic content. Overall the survey on \toolname in its current design phase seem more useful for IT and CS majors, however significant changes are still needed.


\section*{Introduction}
Teaching computer programing and related software design skills can be a challenging task, particularly while working with novice technology students \cite{teodosiev2012some}. Competency in at least one programming language is often part of the minimal requirements for many computing or technology degrees and a prerequisite for higher level courses. It is therefore essential that students have a strong foundation in programming to be successful. Computer programming involves the design and implementation of an algorithm (or a set of algorithms) that solves a problem, using the correct syntax of a particular programming language. Additionally, programs need to be tested and debugged. Teaching not only the correct syntax of the language is challenging, but proper design, form, testing and debugging techniques are also as important.

\section*{Literature Review}
As Gomes and Mendez point out \cite{gomes2007learning}, two of the major difficulties in teaching novice students programming skills is the lack of personalization and emphasis on individual learning style, and that programming languages in general, are meant for professional use in the field and not designed specifically for educational purposes. Online programing courses can pose additional challenges for educators related to student learning style \cite{ccakirouglu2014analyzing}. Research suggests that students benefit from a more student centered active learning approach compared to a more traditional lecture course format \cite{zhang2013teaching}. To overcome some of these difficulties, there have been various approaches in teaching programming skills to improving pedagogy. In some instances, schools have adopted visual programming languages such as Alice and Scratch as a more “gentle” introduction to code. These languages provide a visual interface where students can work and write programs. Chang \cite{chang2014effects} points out with Alice, that “Abstract concepts can be transformed into visual representations that help students develop debugging skills, observe variables, and trace logic, which leads to correcting their programs and solving problems” (pg. 190). These languages are geared not for full application development, but for learning logic, control structures and sometimes object oriented design. Python is increasingly being used in introductory courses as an easier to learn language compared to C++ or Java \cite{shein2015python}. There have been other approaches towards programming visualization or having highly visual programming environments for students such as ViLLE \cite{rajala2007ville}, Jeliot 3 \cite{moreno2004visualizing}, or BlueJ \cite{kolling2003bluej}. There are additional visualization tools that can be helpful for novice programmers to learn the fundamentals.

In addition to changes in language selection and use of visual environments, other approaches tied into pedagogy of teaching programing concepts are being explored. The use of Learning Objects (LO) are increasingly being explored which involves the creation of many exploratory and full scale learning tools \cite{adamchik2003learning,mazilu2015learning}. Many LOs include aspects of animation, annotation, examples and visualization as a modularized component, which can often be integrated or developed for use in a Learning Management system \cite{boyle2003design}. Some of the mentioned visual environments or languages can be used in conjunction with or as part of a LO. Some aspects of learning particular programming constructs can be emphasized as part of a LO. For example, conditional statements, loops, functions, arrays, etc. Loop centric visualization prototypes have been developed, such as by Olsson and Mozelius \cite{olsson2015visualization}. Our approach, while focusing on loops, is fully interactive, web based and includes various forms of feedback including voice over explanations. We envision this project to be an additional aide in the “toolset” that students have to be successful using an active approach to learning to program.

\subsection*{System Usability Scale (SUS)}
Portions of the evaluation survey used to assess the usability of \toolname was based on the System Usability Scale (SUS) \cite{brooke1996sus}. At this phase of evaluation of the tool, it was important to gauge the general usability and potential usefulness of the program. Participants were asked to rank how well they agreed over a continuum between “strongly agree” and “strongly disagree” over a short set of response statements using a Likert scale. SUS provides for a dependable tool to measure reliability of a system based on ten questions, which provide general but useful and valid results \cite{bangor2009determining}. Scoring of results are based on a scale from 0-100 and are used to classify the ease of use of a particular program, environment or website \cite{usabilityGovSite}.

\section*{\toolname}

\toolname allows the user to act as the code compiler and step through each line of code seeing the logic flow and what values get assigned to variables in the program. The tool has two basic modes of operation, demonstration mode and evaluation mode. For research purposes, an example of each was set-up for the users to test the functionality of both modes.

\begin{figure}
\centering
\includegraphics[width=0.9\textwidth]{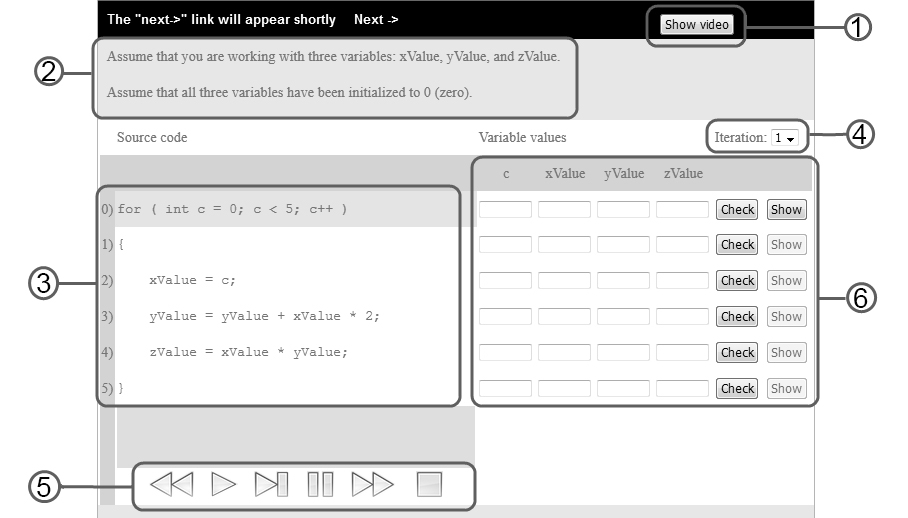}
\caption{\toolname in Demonstration Mode}
\label{fig:1}
\end{figure}

\subsection*{Demonstration Mode}

In demonstration mode, shown in Figure \ref{fig:1}, the page displays several pieces of information to the user. The examples were preceded by a video explanation and demonstration of the tool functionality which the user could toggle on and off with a button at the top of the screen (1). At the top of the page, any initial parameters or assumptions about the code are documented to the user (2). The program source code is shown with numbered lines similar to a typical development environment with the current position highlighted (3). On each line of code, text input boxes are arranged in a grid with each column labeled with a variable from the program (4, worksheet area). The input boxes show the value of the variables as the code executes or allows the user to enter and check the value. Alongside the input boxes, there are two buttons labeled “check” and “show”. The check button uses green and red colors to indicate if the user has entered a correct or incorrect value for that line of code. The show button immediately displays the correct values for the line of code.

Another set of buttons sits directly below the code which represents the functionality of a typical MP3 player (5). Each line of code has an associated audio file which can be played to increase understanding. This allows the user to see the number of loops they have completed. The actions associated with the buttons are the following, left to right: 1) Move Back, 2) Play All, 3) Play Current Line, 4) Pause Audio, 5) Move Forward, and 6) Stop All. For any code that loops, there is a drop-down menu labeled “iteration” (6).

\begin{figure}
\centering
\includegraphics[width=0.9\textwidth]{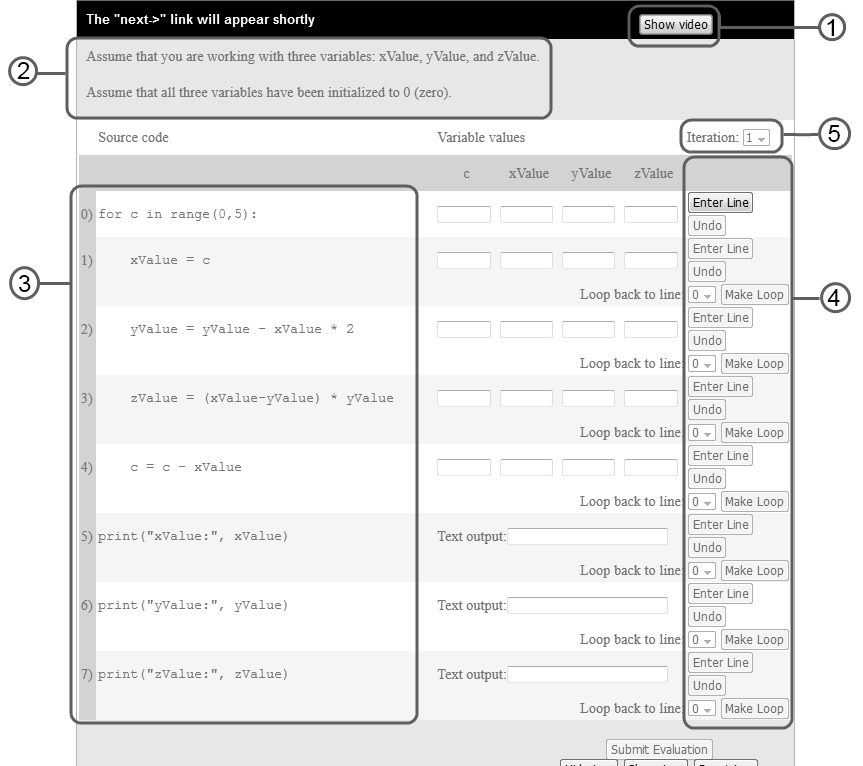}
\caption{\toolname in Evaluation Mode}
\label{fig:2}
\end{figure}

\subsection*{Evaluation Mode}

The evaluation mode, shown in Figure \ref{fig:2}, provides the user the ability to define how the compiler will behave and gives specific functionality for this purpose. This mode has some similarities with the demonstration mode as well as some differences. Another video demonstration preceded the evaluation mode testing with a button to toggle the visibility (1). Like the demonstration mode, the initial parameters are shown (2) as well as numbered lines of code with input boxes (3) allowing the user to enter variable values. Since this is an evaluation, no feedback is given to let the user know if they have entered the correct or incorrect values.

The main difference exists in the worksheet. Once the user enters all the values on a line, they can press the “enter line” button (4) to lock those input boxes and advance to the next line. If the user realizes they have entered an incorrect value, they can press the undo button (4) to go back to the previous line.

When a loop exists in the code, there are specific inputs to allow the user to show what line the compiler jumps to. There is a pull-down input labeled “Loop back to line:” which contains the numbers corresponding to all the previous lines of code. When the user selects the line they wish to jump to, they then click the “Make loop” button to move to that line. This action advances the “iteration” indicator (5) by 1 and clears all the input boxes back to that line while the previously entered values get stored for the final submission. Once all the lines have been completed, the “submit evaluation” button becomes active and the user can send their answers to be graded.

\section*{Profile of the Students}

The questionnaire was administered to students who were taking CS0, CS1, and CS2 courses at three metropolitan institutions in the Greater Baltimore area during the Spring 2016 semester. 129 people completed the survey. Our first goal was to identify the type of degree that the students are working on. Table \ref{table:1} shows a detailed breakdown of degrees.

\begin{table}
\centering
\caption{Frequency of academic programs\label{table:1}}
\begin{tabular}{c|c}
\textit{Academic Program} & \textit{Students} \\
\hline \hline
IT & 46\\
CS & 44\\
IS & 10\\
\hline
Technology Majors & 100\\
\hline \hline
Mathematics \& Physical Science & 22\\
Liberal Arts & 5\\
Business \& Economics & 1\\
Fine Arts & 1\\
\hline
Other Majors & 29\\
\hline \hline
Total & 129\\
\end{tabular}
\end{table}

A total of 100 students (78\%) are enrolled in technology-related majors with 46 in Information Technology (IT), 44 in Computer Science (CS) and 10 in Information Systems (IS). Collectively we refer to these three groups as technology majors, giving the group total as well as the breakdown by degree in the rest of the article.

Among the students who answered the survey, 22 were enrolled in Mathematics and Physical Sciences. We were surprised to see several students (5\%) who were pursuing non-STEM majors enrolled in the programming courses, which included Liberal Arts (5), Business (2), and Fine Arts (1). We will refer to these non-technology-related majors as other majors.

\begin{table}[h]
\parbox{.48\linewidth}{
\centering
\caption{First Programming Course\label{table:2}\\}
\begin{tabular}{c|c|c}
\textit{Academic Program} & \textit{Yes} & \textit{No}\\
\hline \hline
IT & 10 & 34 \\
CS & 9 & 27 \\
IS & 2 & 8 \\
\hline \hline
Technology Majors & 21 & 69 \\
\hline
Other Majors & 21 & 8\\
\hline \hline
Total (129) & 42 & 77
\end{tabular}
}
\hfill
\parbox{.48\linewidth}{
\centering
\caption{Number of Completed Programming Courses\label{table:3}}
\begin{tabular}{c|c|c|c|c|c}
\textit{Acad. Prog.} & \textit{0} & \textit{1} & \textit{2} & \textit{3} & \textit{4+}\\
\hline \hline
IT & 9 & 18 & 11 & 4 & 4 \\
CS & 10 & 18 & 10 & 4 & 2 \\
IS & 2 & 4 & 4 & 0 & 0 \\
\hline \hline
Tech. Majors & 21 & 40 & 25 & 8 & 6 \\
\hline
Other Majors & 21 & 6 & 2 & 0 & 0 \\
\hline \hline
Total (129) & 42 & 46 & 27 & 8 & 6
\end{tabular}
}
\end{table}

We then collected information on their programming experience. Since we are working with students in CS0, CS1, and CS2 we expected a wide difference of experience. First, the majority of students (60\%) were not in their first programming course, as shown in Table \ref{table:2} and Table \ref{table:3}. While 42 students were in their first programming course when completing the survey, 46 students had already completed one course. We also had 35 students who completed two or three courses and 6 students who completed four or more courses.

\begin{table}[h]
\centering
\caption{Programming Experience Level (1: Not experienced, 5: Very experienced) \label{table:4}}
\begin{tabular}{c|ccccc|c}
\textit{Academic Program} & \textit{1} & \textit{2} & \textit{3} & \textit{4} & \textit{5} & $\bar{x}$\\
\hline \hline
IT & 5 & 18 & 17 & 6 & 0 & 2.52 \\
CS & 6 & 14 & 17 & 7 & 0 & 2.57 \\
IS & 1 & 1 & 8 & 0 & 0 & 2.70 \\
\hline \hline
Technology Majors & 12 & 33 & 42 & 13 & 0 & 2.56 \\
\hline
Other Majors & 11 & 9 & 7 & 2 & 0 & 2.00\\
\hline \hline
Total (129) & 23 & 42 & 49 & 15 & 0 & 2.43
\end{tabular}
\end{table}

In the next few questions we asked the students to quantify their answer on a Likert scale ranging from 1 to 5. First, we wanted to look at their self-confidence with programming by asking about their experience, comfort level, and attitude towards this activity. We asked the students to rate their experience, ranging from not experienced at all (1) to very experienced (5). Among the results, reported in Table \ref{table:4}, we could see that IT students (2.52) regard themselves as the least experienced among the technology majors (2.56). As expected, the mean for the responses of other majors (2.00) was lower than that of technology majors.

\begin{table}[h]
\centering
\caption{Comfort Level with Programming (1: Not comfortable, 5: Very comfortable) \label{table:5}}
\begin{tabular}{c|ccccc|c}
\textit{Academic Program} & \textit{1} & \textit{2} & \textit{3} & \textit{4} & \textit{5} & $\bar{x}$\\
\hline \hline
IT & 9 & 12 & 17 & 7 & 1 & 2.54 \\
CS & 1 & 10 & 17 & 12 & 4 & 3.18 \\
IS & 1 & 3 & 4 & 2 & 0 & 2.70 \\
\hline \hline
Technology Majors & 11 & 25 & 38 & 21 & 5 & 2.84 \\
\hline
Other Majors & 3 & 10 & 12 & 3 & 5 & 2.91\\
\hline \hline
Total (129) & 14 & 35 & 50 & 24 & 10 & 2.86
\end{tabular}
\end{table}

Next we asked the students about their comfort level with programming, and the results are reported in Table \ref{table:5}. We asked the students to rate their comfort level, ranging from not comfortable at all (1) to very comfortable (5). Like experience level, we see the mean reported comfort level for IT majors (2.54) is the lowest among all the technology majors (2.84). Interestingly, other majors reported a higher comfort level (2.91) than IS (2.70) or IT (2.54) students. Generally, students rated their comfort level (2.86) higher than their experience level (2.43).

\begin{table}[h]
\centering
\caption{Attitude towards Programming (1: Hate it, 5: Love it) \label{table:6}}
\begin{tabular}{c|ccccc|c}
\textit{Academic Program} & \textit{1} & \textit{2} & \textit{3} & \textit{4} & \textit{5} & $\bar{x}$\\
\hline \hline
IT & 7 & 10 & 14 & 11 & 4 & 2.89 \\
CS & 0 & 4 & 4 & 20 & 16 & 4.09 \\
IS & 2 & 3 & 3 & 2 & 0 & 2.50 \\
\hline \hline
Technology Majors & 9 & 17 & 21 & 33 & 20 & 3.38 \\
\hline
Other Majors & 6 & 2 & 10 & 8 & 3 & 3.00\\
\hline \hline
Total (129) & 15 & 19 & 31 & 41 & 23 & 3.29
\end{tabular}
\end{table}

We then went on to assess their attitude towards programming, reported in Table \ref{table:6}. We asked attitude towards programming from I hate it (1) to I love it (5). This time, IS students (2.50) reported the least favorable feelings towards programming among technology majors (3.38). As expected, CS majors (4.09) reported the highest mean. Interestingly, IS majors reported the least favorable feelings towards programming compared to students of other majors (3.00) and the whole group (3.29).

\begin{table}[h]
\centering
\caption{Attitude towards Programming Courses (1: Not helpful at all, 5: Very helpful)\label{table:7}}
\begin{tabular}{c|ccccc|c}
\textit{Academic Program} & \textit{1} & \textit{2} & \textit{3} & \textit{4} & \textit{5} & $\bar{x}$\\
\hline \hline
IT & 5 & 9 & 14 & 12 & 6 & 3.11 \\
CS & 1 & 7 & 10 & 16 & 10 & 3.61 \\
IS & 0 & 3 & 5 & 2 & 0 & 2.90 \\
\hline \hline
Technology Majors & 6 & 19 & 29 & 30 & 16 & 3.31 \\
\hline
Other Majors & 5 & 7 & 7 & 5 & 5 & 2.93\\
\hline \hline
Total (129) & 11 & 26 & 36 & 35 & 21 & 3.22
\end{tabular}
\end{table}

Next, we wanted to assess the students' attitude towards the helpfulness of past and present programming courses. For this question, the students were asked to rate their experience with such courses from did not help at all (1) to helped very much (5). The results are reported in Table \ref{table:7}. In this case, technology majors differed on their reported attitude towards programming courses. Clearly CS students found programming courses most helpful (3.61), followed by IT students with a mean of 3.11. Interestingly, other majors (2.93) had a slightly higher mean attitude than IS students (2.90).

\begin{table}[h]
\centering
\caption{Students who have used the Internet to learn how to program\label{table:8}}
\begin{tabular}{c|c|c}
\textit{Academic Program} & \textit{Yes} & \textit{No}\\
\hline \hline
IT & 39 & 7 \\
CS & 33 & 11 \\
IS & 6 & 4 \\
\hline \hline
Technology Majors & 78 & 22 \\
\hline
Other Majors & 14 & 15\\
\hline \hline
Total (129) & 92 & 37
\end{tabular}
\end{table}

Lastly we asked students whether they used the Internet to learn how to program. The results are reported in Table \ref{table:8}. While 71\% of all students reported having used the Internet to supplement learning how to program, it varied substantially by discipline. Three-quarters (74\%) of technology majors reported using the Internet to supplement learning how to program. Just under half (48\%) of other majors reported using the Internet.

Students were asked, in an open-ended question, what Internet resources they were already using to help learn programming. Results are reported in Table \ref{table:9}. Of the 129 students completing the questionnaire, 92 reported using Internet resources while 37 indicated they have not.

\begin{table}[h]
\centering
\caption{Internet resources used by students to learn how to program\label{table:9}}
\begin{tabular}{c|cccc}
\textit{Acad. Program} & \textit{YouTube Videos} & \textit{Prog. Websites} & \textit{Google} & \textit{Online Forums}\\
\hline \hline
IT & 21 & 21 & 7 & 2 \\
CS & 15 & 13 & 3 & 0 \\
IS & 3 & 5 & 1 & 0 \\
\hline \hline
Tech. Majors & 39 & 39 & 11 & 7 \\
\hline
Other Majors & 5 & 4 & 5 & 3\\
\hline \hline
Total (129) & 44 & 43 & 16 & 10
\end{tabular}
\end{table}

YouTube was the number one (34\%) Internet resource used by students to learn programming. The second most popular resource were programming websites (33\%) such as StackOverflow, Codecademy, and W3Schools. Interestingly, Google was the third most popular response for online resources (12\%), suggesting some students did not make a distinction between the search engine and the actual content providers. Online forums and discussion boards were just behind Google as the fourth most popular online resource (8\%). Students referenced “coding” and “programming” forums without providing specific forum names. Finally, a few students reported accessing online documentation from vendors such as Microsoft and Oracle.

\section*{Results and Discussion}

After the students completed the demographic section, they were presented with the two pages described earlier. After the students completed the Demo mode, they were allowed to complete the SUS questionnaire regarding that section. After that, they were asked to go through the Evaluation mode, and then they took another SUS questionnaire regarding this second mode.

In this section we will focus solely on students in the IT, CS, and IS programs. The results for IT and CS students are most significant, because we had 46 and 44 respondents respectively. The pool of users included only 10 IS students. Although the SUS results can range from 0 to 100, we are most interested in the relative difference between groups for each mode, and the differences within the majors. Looking at the absolute values, we can approximately identify the boundaries reported in Table \ref{table:10} \cite{gomes2007learning}. These values are particularly important because usability is directly correlated to the success of e-learning tools \cite{melis2003lessons}.

\begin{table}[h]
\centering
\caption{Classifications related to SUS scores\label{table:10}}
\begin{tabular}{c|c}
\textit{Adjective} & \textit{SUS Score}\\
\hline \hline
Worst imaginable & From 0 to 25 \\
Poor & From 25 to 38 \\
OK & From 38 to 52 \\
Good & From 52 to 73 \\
Excellent & From 73 to 85 \\
Best imaginable & From 85 to 100 \\
\end{tabular}
\end{table}

The overall SUS results are shown in Table \ref{table:11}. Overall IT students found \toolname most usable for both the narrated demonstration as well as the evaluation systems. IS students found the tools the least usable. As we mentioned earlier, we can expect a correlation between usability and the potential success of an e-learning tool, so we believe that \toolname has great potential in IT courses.

\begin{table}[h]
\centering
\caption{SUS results by academic program\label{table:11}}
\begin{tabular}{c|c|c}
\textit{Academic Program} & \textit{Narrated demo} & \textit{Evaluation} \\
\hline \hline
IT & 60.1 & 59.8 \\
\hline
CS & 57.2 & 56.3 \\
\hline
IS & 49.5 & 52.2
\end{tabular}
\end{table}

When we look at the average SUS values differentiating between students for whom this was their first programming course, reported in Table \ref{table:12}. IT students find the tool usable enough for both the narrated demonstration (N) as well as the evaluation (E) component. CS students for whom this is the first programming course report the tools to be more usable, and consequently they may be more prone to using it as supplemental material for their studies. IS students who are new to programming do not seem to find this tool usable at all.

\begin{table}[h]
\centering
\caption{Results by first programming course (N: Narrated demo, E: Evaluation)\label{table:12}}
\begin{tabular}{c|c|c|c}
\textit{Academic Program} & \textit{First course} & \textit{Not first course} & \textit{Mode} \\
\hline \hline
IT & 58.3 & 60.5 & N\\
& 62.8 & 59.1 & E\\
\hline
CS & 60.2 & 56.3 & N\\
& 61.0 & 54.9 & E\\
\hline
IS & 31.2 & 54.1 & N\\
& 40.0 & 55.3 & E
\end{tabular}
\end{table}

In Table \ref{table:13} we can see that most IT students who feel comfortable with programming (1=least comfortable, 5-most comfortable) find \toolname usable. CS students who are not comfortable with programming do not find it as usable. IS students who feel neutral about their comfort with programming find this tool to be most usable. No significant trend was identifiable in the breakdown of SUS values for this category.

\begin{table}[h]
\centering
\caption{SUS results by Comfort Level (1: low, 5: high)\label{table:13}}
\begin{tabular}{c|ccccc|c}
\textit{Academic Program} & \textit{1} & \textit{2} & \textit{3} & \textit{4} & \textit{5} & \textit{Mode}\\
\hline \hline
IT & 63.3 & 61.0 & 54.4 & 69.6 & 47.5 & N \\
& 56.9 & 61.0 & 57.3 & 70.0 & 40.0 & E \\
\hline
CS & 32.5 & 57.7 & 60.6 & 50.2 & 68.7 & N \\
& 32.5 & 53.2 & 59.3 & 53.5 & 65.6 & E \\
\hline
IS & 32.5 & 37.5 & 62.5 & 50.0 & --- & N \\
& 37.5 & 39.2 & 68.1 & 47.5 & --- & E
\end{tabular}
\end{table}

If we look at the mean SUS value by how the students feel about programming courses (1-not helpful, 5-very helpful), we can see an interesting trend, reported in Table \ref{table:14}. IT students who find programming courses less useful find the tool most usable, while CS students in the same situation feel in the opposite way. It appears that \toolname receives the best reviews from students who feel somewhat neutral about the usefulness of programming courses, leading us to think that this particular population may benefit the most from tools such as this, with this particular design.

\begin{table}[h]
\centering
\caption{SUS results by Reaction to Programming Courses (1: low, 5: high)\label{table:14}}
\begin{tabular}{c|ccccc|c}
\textit{Academic Program} & \textit{1} & \textit{2} & \textit{3} & \textit{4} & \textit{5} & \textit{Mode}\\
\hline \hline
IT & 73.0 & 48.1 & 62.0 & 61.7 & 59.6 & N \\
& 67.0 & 50.3 & 62.7 & 61.0 & 58.7 & E \\
\hline
CS & 32.5 & 53.9 & 60.5 & 54.5 & 63.0 & N \\
& 32.5 & 48.6 & 62.0 & 52.7 & 64.2 & E \\
\hline
IS & --- & 39.2 & 53.0 & 56.2 & --- & N \\
& --- & 49.2 & 52.0 & 57.5 & --- & E
\end{tabular}
\end{table}

Overall the current design of \toolname is a good first step. Although this is not the first iteration of this tool, this is the first time we have a system that students can use for interpreting code. We believe that this feature has great potential and that students will utilize it, with some improvements in the User Interface.

We believe that one significant drawback of this current design is not attributable just to the interface. The examples that were used for this experiment involved loops, and not all students are able to easily track the value of different variables as the loop goes on. If we think of the value of variables in a loop as a three-dimensional structure where each individual variable has a different value depending on the line within the loop as well as the iteration number of the loop, we are basically trying to portray a 3-dimensional worksheet in a 2-dimensional HTML page. The worksheet area that students can use to keep track of the value of variables must be improved to also accommodate looping more easily. We believe that giving students examples that do not include looping at first may improve the usability of the tool. On the other hand, we should design the new interface with looping in mind, to offer a consistent interface for examples that deal with conditions, functions, and looping. We will also explore different types of worksheets, where students are not overwhelmed by a series of input areas that they need to complete.

Currently the system is quite bulky, especially with all the controls for the audio playback. It is important that we integrate those controls with the interface in a way that is more streamlined. The students seemed to struggle finding their way through the commands. We will integrate such commands with context-sensitive menus that only appear when students hover over a particular line, for example.

\section*{Conclusions}

Overall the results were very promising. This first iteration of a media-rich \toolname has shown great potential as supplemental instruction material for computer programming courses, especially among IT students. We will proceed with improvements to the user interface, which will be followed by more testing. At that point, once the tool is more streamlined and with easy interaction, we will conduct testing on whether this system can make a difference or not in the classroom if used as an integral part of the instructional process.

\nolinenumbers

\bibliography{splashe}

\bibliographystyle{abbrv}

\end{document}